\documentclass[]{raa}

\usepackage{natbib}
\usepackage{amssymb,amsmath}
\usepackage{graphicx,times}
\bibpunct{(}{)}{;}{a}{}{,}
\usepackage[pdftex,a4paper=true,pagebackref=true]{hyperref}
\hypersetup{colorlinks = true, linkcolor = green, anchorcolor = red, citecolor = blue, filecolor = red, pagecolor = red, urlcolor = red}

\begin{document}
    \title{Search for strong galaxy-galaxy lensing in SDSS-III BOSS}
    \volnopage{Vol.0 (20xx) No.0, 000--000}      
    \setcounter{page}{1}          
    \author{Xinlun Cheng
        \inst{}\footnote{Based on observations obtained with MegaPrime/MegaCam, a joint project of CFHT and CEA/DAPNIA, at the Canada-France-Hawaii Telescope (CFHT) which is operated by the National Research Council (NRC) of Canada, the Institut National des Sciences de l'Univers of the Centre National de la Recherche Scientifique of France, and the University of Hawaii. Based on spectroscopic data from the Baryon Oscillation SpectroscopicSurvey of the Sloan Digital Sky Survey III.}}
    \institute{Department of Astronomy, Tsinghua University, Beijing 100084, China; {\it chengxinlun@gmail.com}\\
        \vs\no
        {\small Received  20xx month day; accepted  20xx  month day}}

    \abstract{Strong lensing is one of the most spectacular views in the universe. Many cosmological applications have been proposed, but the number of such lensing systems is still limited. In this work, we applied an improved version of a previously developed spectroscopic lensing search method to the SDSS-III BOSS and proposed a list of highly possible candidates. Follow-up CFHT Megacam imaging observations were performed for five systems, and two out of five are probably strong lensing systems with at least one image close to the central galaxy, although no counter images are detected.
\keywords{gravitational lensing: strong --- techniques: spectroscopic --- methods: data analysis}}

    \authorrunning{X. Cheng}
    \titlerunning{Strong galaxy-galaxy lensing in SDSS-III BOSS}

    \maketitle

    \section{Introduction}
    Strong galaxy-galaxy lensing produces multiple, distorted images of a background galaxy through the gravitational field of the foreground galaxy. It was first suggested by \citet{zwicky} for galaxy-clusters, and the first discovery of a galaxy-scale lens was made by \citet{walsh}. The development in optics has led to a rapid increase in the number of strong lensing system discovered. As in 2010, the total number of confirmed strong gravitational lensing by galaxies was approximately 200 \citep{sl_review}.
    
    A number of methods have been utilized to discover strong gravitational lensing by galaxies. These include manual or automatic inspection of high-resolution images from optical or radio telescopes \citep{1996ApJ...467L..73H, 1999AJ....117.2010R, 2003MNRAS.341...13B, 2004ApJ...600L.155F, 2007ApJ...660L..31M, 2008ApJS..176...19F, 2008MNRAS.389.1311J, 2009ApJ...694..924M, 2017MNRAS.468.3757W}, employing citizen science \citep{spacewarp} or machine learning \citep{subpca, 2017A&A...597A.135B, 2017MNRAS.471..167J, 2017MNRAS.472.1129P, 2018MNRAS.473.3895L} approaches in sky surveys, detection of emission lines in optical spectroscopy with emission from foreground galaxy removed \citep{1996MNRAS.278..139W, slacs, 2006AJ....132..999O, 2008AJ....135..512O, bells, 2017ApJ...851...48S}, examination of 2D distribution of redshift in a small region with integral field unit (IFU) or high-resolution spectrograph \citep{1997ApJ...474L...1Z, 2007NJPh....9..443B, 2012ApJ...758L..17B}, and detection of variability in time-domain \citep{2005ApJ...626..649P}. Compared to other methods, spectroscopic methods have the advantage of being able to obtain the spectroscopic redshift of both the lens and the source galaxies and the velocity dispersion of the lens galaxy, but lensing system discovered by these methods often require imaging confirmations from optical telescopes.

    Many physical implementations could be found in strong lensing by galaxies. Strong gravitational lensing is an indispensable tool in the measurement of the projected mass distributions of galaxies, including both dark and luminous matter \citep{1937ApJ....86..217Z, 2004ApJ...611..739T, 2014MNRAS.439.2494O, 2018MNRAS.480..431L}. By analyzing the fine structures of an Einstein ring, one could study the gravitational perturbation of unseen dark matter substructure, which could reveal the distribution of dark matter halos and distinguish cold dark matter (CDM) model from warm dark matter (WDM) model \citep{2001ApJ...561...46K, 2002ApJ...572...25D, 2013MNRAS.435L..53P}. If a point source with variable flux under constant $\rm H_0$, such as a quasar, is gravitationally lensed, it can put constraint on cosmological parameters, especially the expansion rate $H(z)$, by the lens time delay measurement \citep{2005IAUS..225..297C, 2013A&A...556A..22T, 2013ApJ...766...70S}. Recent measurements are capable of limiting the Hubble Constant $H_0$ to 3.8\% precision \citep{2017MNRAS.465.4914B}. Since strong lensing always magnifies the source objects, it could be used as cosmic telescope, revealing details of distant galaxies and quasars. It can provide estimation for the mass of super massive black holes for galaxies, which is one of the few methods to estimate the mass of centeral black holes of non-AGN galaxies \citep{2001MNRAS.323..301M}. Based on the variability of source images, \citet{2008ApJ...673...34P} shed light on the spatial structure of the accretion disk of a quasar. \citet{2013ApJ...777L..17V} revealed a star-forming galaxy at z=3.42 with the help of a lens galaxy at z=1.53. While supernova is difficult to detect beyond z=1.5, gravitational lensed supernova have been discovered \citep{2016ApJ...820...50R, 2017Sci...356..291G} and offered an unprecedented view of the high-redshift universe.

    The Sloan Digital Sky Survey III (SDSS-III) \citep{2011AJ....142...72E} initiated the Baryon Oscillation Spectroscopic Survey (BOSS), and with several upgrades to the original equipment, approximately 1.5 million spectra for luminous galaxies with redshift $z<1.0$ were obtained. The upgrades include reducing the fiber diameter to 2 arcseconds, increasing efficiency in grating and CCD \citep{2013AJ....145...10D}. This provides us with a wonderful opportunity to search for strong galaxy-galaxy lensing systems using spectroscopic methods.
    
    Although fully automatic strong lensing detection using imaging data has been extensively researched \citep{2007A&A...472..341S, 2014ApJ...785..144G, 2019A&A...625A.119M}, such attempt in spectroscopic detection is rare partially due to the lack of data as training set for an automatic detection program. Recently, several highly-possible catalogs have been composed and confirmed by images from Hubble Space Telescope (HST) \citep{slacs, bells}. In this work, we applied a similar method to the entire SDSS-III BOSS sample, fully released in SDSS DR12 \citep{2015ApJS..219...12A}, and produced a catalog of possible strong lensing candidates, in which we confirmed two out of five candidates with CFHT Megacam as a pilot study.

    The outline of the paper is as following. In Section \ref{sec:algo}, we describe our selection and contamination removal algorithm. Results are reported in Section \ref{sec:res}, and the processing and analysis of CFHT Megacam images are in Section \ref{sec:cfht}. We discuss our finding in Section \ref{sec:dis}. In Section \ref{sec:con}, we present our conclusion.
    
    \section{Selection algorithm}\label{sec:algo}
    \subsection{Algorithm design overview}
    The kernel of spectroscopic detection is that since both the photons from the lens and the background galaxy are captured in the same fiber, the spectrum from that fiber should be the superposition of two spectra of galaxies with different redshifts. In the BOSS spectroscopic pipe line, which is described in \citet{2012AJ....144..144B}, redshift and classification templates for galaxies, quasars, and star classes are constructed by performing a rest-frame principal-component analysis (PCA) of training samples with known redshift. The leading eigenvectors from the PCA are used to define a linear template basis that is used to model the spectra in the redshift analysis. A range of trial galaxy redshifts is explored from redshift -0.01 to 1.00. The combination of redshift and template class that yields the lowest reduced chi-squared is adopted as the pipeline measurement of the redshift and classification of the spectrum. This best fit is then subtracted from the original spectra and yield the reduced spectra for each galaxy, thus removing the foreground emissions from the lens galaxy. However, emission lines from the source galaxy are not removed and should be detected in the reduced spectra.

    The selection algorithm can be divided into two separated procedures: primary selection and post-processing. The primary selection is aimed to select objects with features of strong lensing using information from the corresponding fiber and its adjacent fibers, and the post-processing is mainly built to remove false positives produced by various contamination.
    
    \subsection{Primary selection}
    The basic design of the primary selection is similar to that in \cite{bells}, but with a few tweaks and improvements. The steps are designed as follow:

    \begin{enumerate}
        \item A selection on the type and class of objects is performed to remove QSOs and stars from the sample since we are only interested in galaxy-galaxy lensing.
        
        \item Galaxies whose redshift could not be determined ($\rm zwarning \neq 0$) are removed.
        
        \item Spectra glitches and strong sky emission lines are masked. Residual spectra with unusual high SNR are removed due to possible incomplete template subtraction. A list of all masked lines is provided in \autoref{tab.sky_sdss}.
        
        \item Peak search is performed by maximum-likelihood estimator of a Gaussian fitted to the data \citep{slacs} with $A=1.2$ and $\sigma=30km/s$. Peaks with $\rm SNR > 6.0$ are accepted.
        
        \item A single Gaussian, as well as double Gaussians with same width are fit at each accepted peak, and their reduced chi-square were compared. The one with smaller value is stored.
        
        \item Peaks with wavelength $\lambda>9200\AA$ are rejected to remove contamination from sky emission lines.
        
        \item Remaining peaks are ranked according to the stored chi-square and only the smallest five ones are kept if there are more. If no doublet remains, the object is labeled as non-lensing system. Otherwise move on to the next step.
        
        \item Emission from adjacent fibers are compared and labeled as non-lensing if contamination is confirmed. See the next paragraph for details.
        
        \item Assuming the detected peak is the [OII]$\lambda$3727,3729 doublet, redshift of the source ($z_s$) is inferred and if the difference between that of the lens ($z_l$) is smaller than 0.05, the object is labeled as non-lensing. Otherwise move onto the next step.
        
        \item Supposed positions of other strong emission lines ([H$\beta$]$\lambda$4858, [OIII]$\lambda$4959 and [OIII]$\lambda$5008) are computed. Peak search similar to Step IV is performed to find the exact position of the emission lines. Emission lines are fitted with a single Gaussian and their SNR are calculated and stored for later usage. Only $\rm SNR > 3.0$ is counted as detection. Note that we do not assume a fixed ratio between the flux of [OIII]$\lambda$4959 and that of [OIII]$\lambda$5008.
        
        \item Residual spectra, including emission line fittings, are plotted for easier inspection and the object is labeled as candidate.
        
        \item Repeat the steps until all plates in BOSS has been checked.
    \end{enumerate}

    \begin{table}[h]
        \centering
        \caption{Masked sky emission lines in Primary selection. These are also the prominent night sky emission lines listed in SDSS DR12 documentation.}\label{tab.sky_sdss}
        \begin{tabular}{c c}
            \hline\hline
            Central wavelength (\AA) & Maksed Width (\AA) \\
            \hline
            5577.35 & 20 \\
            6300.31 & 20 \\
            6363.10 & 20 \\
            8344.61 & 20 \\
            9375.98 & 20 \\
            \hline\hline
        \end{tabular}
    \end{table}

    In the BOSS spectrometer setup, photons coming from each fiber are stored in different pixels on the CCD. However, a strong emission line from one fiber might affect nearby pixels, causing a spike, which would be mistaken as a peak by the peak detection program, at the same wavelength in the observer frame. Step VIII is designed to minimize the possibility of such false detection. Reduced spectra from the previous and next fiber are extracted and double Gaussians are fit at the same position of the peak in the current fiber. The ratio of flux of emission lines between current one and adjacent ones are computed, and plotted. A number of randomly selected spectra are examined by eye and classified into two categories: contaminated by nearby fibers or not. We decided that a cut at $x + y < 0.6$, where $x$ is the ratio of flux between the current fiber and the previous one and $y$ is that of the next one, is appropriate since we can get rid of $>95\%$ the contamination while still achieving a high completeness.
    
    \subsection{Post-processing}
    Post-processing is traditionally done by human, but due to the size of our data set, it is necessary to automate the process. Analyzing the candidates that pass through the primary selection steps described above, we found out that there are mainly three sources of false positives: sky emission lines and bad foreground template subtraction, low-z objects, and faint objects at given redshift bin. Here are the steps of our post-processing.
    \begin{enumerate}
        \item Remove candidates with detected $[\rm OII]$ doublets falling within over dense regions in observer frame or foreground rest frame. See \ref{alg:sky} for detail.
        
        \item Remove candidates with redshift of lens galaxy $z_l<0.1$ and I-band magnitude $\rm I > 22$. See \ref{alg:mag} for detail.
        
        \item The sum of the SNR of [OII]$\lambda$3727,3729 doublet, [H$\beta$]$\lambda$4858, [OIII]$\lambda$4959 and [OIII]$\lambda$5008 is calculated and saved as a total SN ratio. Remaining candidates are ranked in decreasing order.
        
        \item Candidates with large relative error ($>20\%$) of velocity dispersion are removed from the list. Candidates with $\sigma_v = 0$ or $850\rm\ km\ s^{-1}$ are also removed, since these two values indicate problematic fitting of velocity dispersion in the pipeline of BOSS.

        \item Candidates with estimated Einstein radius $R_E<0.5''$ are removed due to high ($>90\%$) false positive rate reported by \citet{2008ApJ...682..964B}
      
        \item Remove all previously confirmed lensing system from the list (see \ref{sec:res} for more discussions), and save the first 100 candidates into a highly possible candidate list. Visual inspections of template subtraction are preformed on them to ensure that no bad template fitting occurred in BOSS spectroscopic pipe line.
    \end{enumerate}
    
    \subsubsection{Sky emission and residual emission from foreground}\label{alg:sky}
\begin{figure}
        \centering
        \includegraphics[width=\columnwidth]{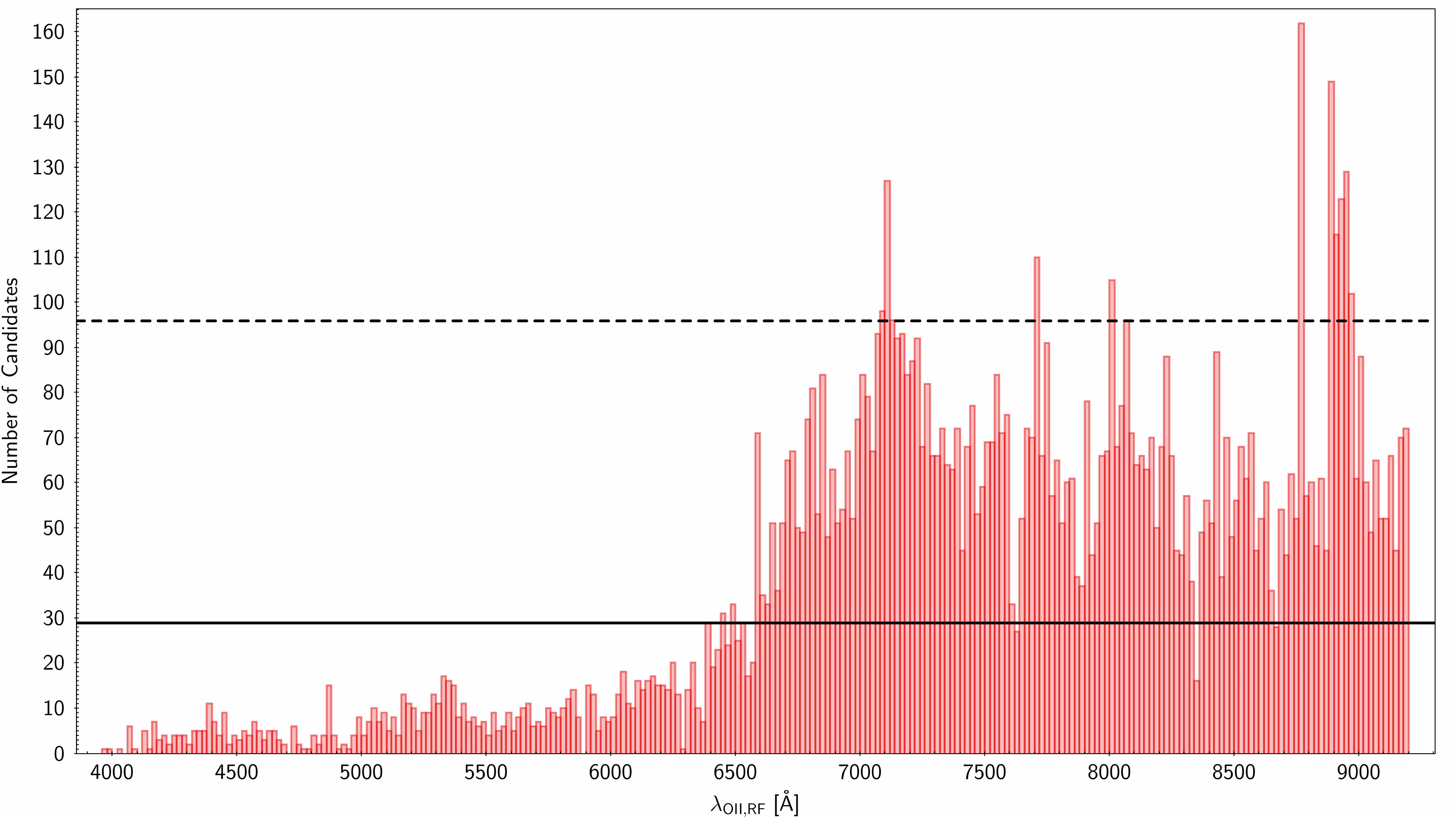}
        \caption{Histogram of wavelength of detected OII doublet in the observer rest frame. False positives caused by sky emission lines will appear as a higher than normal amount of objects in the bin. Solid line indicates the median number of objects in each bin, and the dash line indicates $2\sigma$ confidence level of number of objects in each bin.}\label{fig:o2_orf}
    \end{figure}

\begin{figure}
        \centering
        \includegraphics[width=\columnwidth]{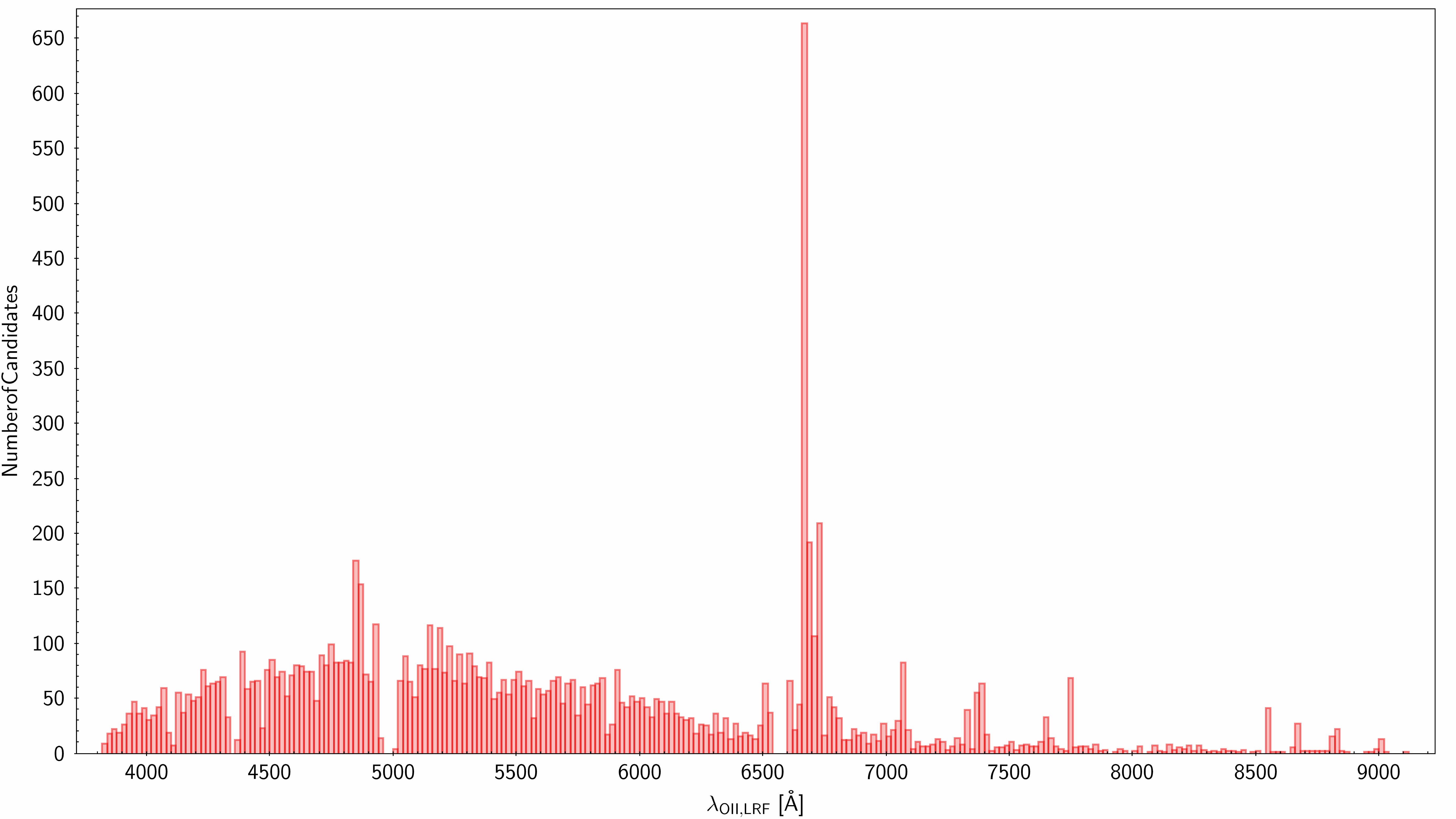}
        \caption{Histogram of wavelength of detected OII doublet in the lens galaxy rest frame. False positives caused by sky emission lines will appear as a higher than normal amount of objects in the bin. Notice a huge rise between 6600\AA and 6740\AA. The possible cause is explained in \ref{alg:sky}}\label{fig:o2_lrf}
    \end{figure}

    Sky emission lines and residual emission lines from foreground galaxies would cause false detection due to their strong intensity. Although it is hard to remove them with information from a few fibers, they will appear as overdense regions when looking at the histogram of the wavelength of the OII doublets of these candidates in observer rest frame (ORF, \autoref{fig:o2_orf}) and lens galaxy rest frame (LRF, \autoref{fig:o2_lrf}).

    In the observer rest frame, the median number of candidates per bin $\rm med=29$ (solid line)
    with standard deviation $\sigma=33.5$. Therefore, we remove candidates whose wavelength of OII doublet lies in bins with number of candidates $>\rm med+2\sigma=96$ (dash line). The corresponding wavelength range is listed in \autoref{tab:sky}.
    \begin{table}
        \centering
        \caption{Contamination removal wavelength range in Post-processing. For candidates with OII doublet wavelength falling into any range listed in this table, they are considered false positives and removed from the candidate list. ORF stands for observer rest frame, and LRF stands for lens galaxy rest frame.}\label{tab:sky}
        \begin{tabular}{c c c}
            \hline\hline
            $\lambda_{\rm start}$(\AA) & $\lambda_{\rm End}$(\AA) & Reason\\
            \hline
            7080 & 7120 & Over dense in ORF \\
            7700 & 7720 & Over dense in ORF \\
            8000 & 8020 & Over dense in ORF \\
            8760 & 8780 & Over dense in ORF \\
            8880 & 8980 & Over dense in ORF \\
            6600(1+$z_l$) & 6740(1+$z_l$) & Over dense in LRF \\
            \hline\hline
        \end{tabular}
    \end{table}

    In the lens galaxy rest frame, contamination removal is performed by comparing over dense regions with a list of galaxy emission lines. We noticed an extreme dense region between 6660\AA\,and 6740\AA. We believe that two reasons might lead to this phenomenon. First, emission lines [SII]$\lambda$6716 and [SII]$\lambda$6730 are within the region and are likely to be falsely detected as source galaxy OII emission line. Second, some galaxies have really strong and wide [H$\alpha$]$\lambda$6562 emission and the BOSS pipe line might have trouble fitting and subtracting it completely, so the leftover emission might be picked up by the Primary Selection program and identified as OII doublet of the source galaxy. The corresponding wavelength range is listed in \autoref{tab:sky}.
    

    \subsubsection{Low redshift objects and Faint objects}\label{alg:mag}
\begin{figure}
        \centering
        \includegraphics[width=\columnwidth]{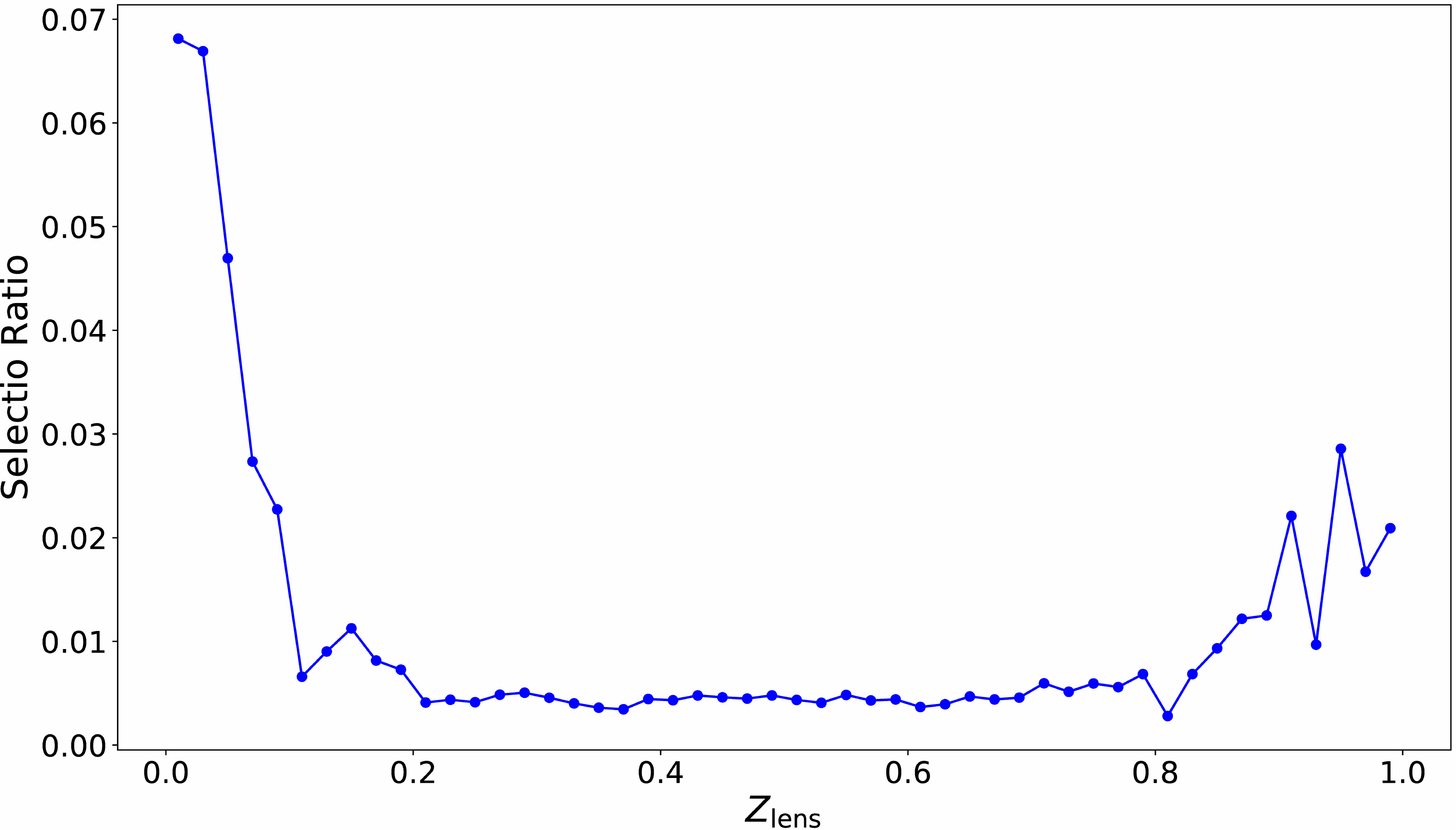}
        \caption{Redshift of lens galaxy ($z_l$) versus selection ratio. Unusually high selection ratio could be observed at $z<0.1$. Therefore, objects with lens galaxy redshift less than 0.1 are much more likely to be false positives and are removed from the candidate list.}\label{fig:z}
    \end{figure}

    A number of objects targeted by BOSS have redshift $z<0.1$, which are unlikely to become strong lensing systems. However, these objects tend to be bright and since the template could not cover every small emission lines, they take up a large proportion of candidates. Therefore, all candidates with redshift $z<0.1$ are removed in the last step. This could be confirmed by plotting redshift $z$ versus selection ratio (\autoref{fig:z}).  The rise in the selection ratio at $z>0.8$ is due to the limited number of objects in high redshift. A similar cut in redshift can also be found in both SLACS \citep{slacs} and BELLS \citep{bells}.
    \begin{figure}
        \centering
        \includegraphics[width=\columnwidth]{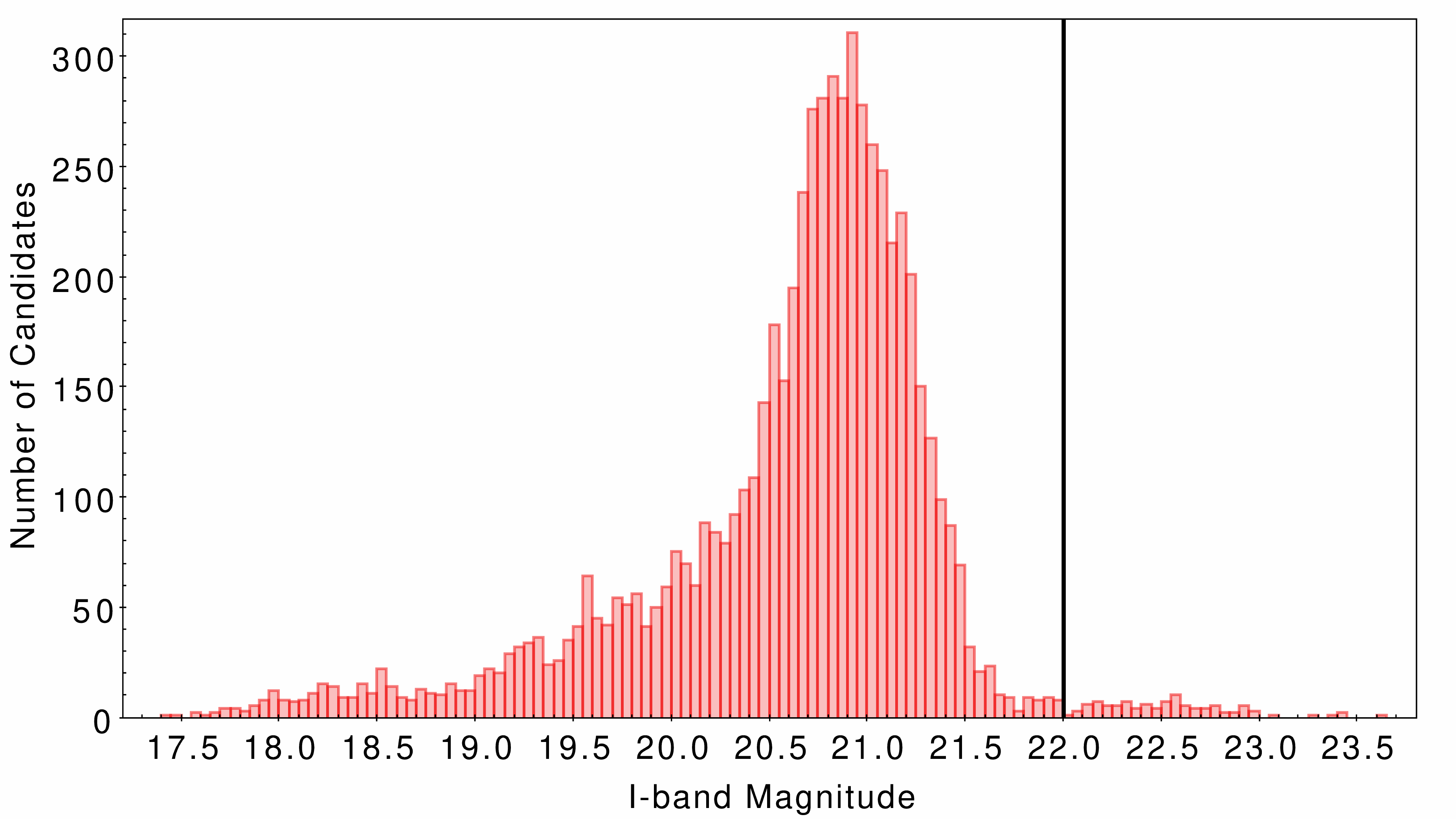}
        \caption{Histogram of I-band magnitude after extinction correction. Faint objects are usually not massive and thus have a lower possibility of gravitational lens the source galaxy.}\label{fig:mag}
    \end{figure}

    Faint objects are usually believed to be less massive when compared to bright objects with similar redshift. Therefore, it is less likely for faint foreground objects to act as lens. From \autoref{fig:mag}, we empirically removed candidates with I-band magnitude after extinction correction $\rm magI - 1.698 ebv > 22$, where $ebv$ is the extinction correction. This threshold is chosen due to two reasons. First, there is a sharp drop in the number of candidates, and secondly, as presented by SLACS and BELLS, no candidate with I-band magnitude fainter than 22 is selected.

    \section{Selected Candidates}\label{sec:res}
    A total number of 2388 plates are in SDSS BOSS, and in each plate there are 1000 fibers. Target selection by object type and class returns 1500002 galaxies. \citet{bells} have already checked several hundred plates, so these plates are excluded from counting the number of candidates. Thus we would only run the selection algorithm on 1380002 galaxies. After running through the steps in Section \ref{sec:algo}, the number of candidates is 396. The detailed numbers of candidates remaining and left after each step are presented in \autoref{tab.boss}.
    \begin{table}
        \centering
        \caption{Number of candidates remaining and rejected after each key step described in Section \ref{sec:algo}}\label{tab.boss}
            \begin{tabular}{c c c}
                \hline\hline
                Key step & Candidates rejected & Candidates left \\
                \hline
                Unusually high spectra SNR & 12806 & 1367196 \\
                Peak search & 1250086 & 117110 \\
                Singlet fit better & 72833 & 44277 \\
                $\lambda>9200$ & 20711 & 23566 \\
                Adjacent fiber contamination & 4114 & 19452 \\
                $z_s - z_l < 0.05$ & 9627 & 9825 \\  
                Sky and bad subtraction & 2495 & 7330 \\
                Low redshift & 931 & 6399 \\
                Faint objects & 116 & 6283 \\
                Other emission lines & 4753 & 1530 \\
                Inaccurate velocity dispersion measurement and $R_E<0.5''$ & 1134 & 396 \\
                \hline\hline
            \end{tabular}
    \end{table}
    
\begin{figure}
        \centering
        \includegraphics[width=\columnwidth]{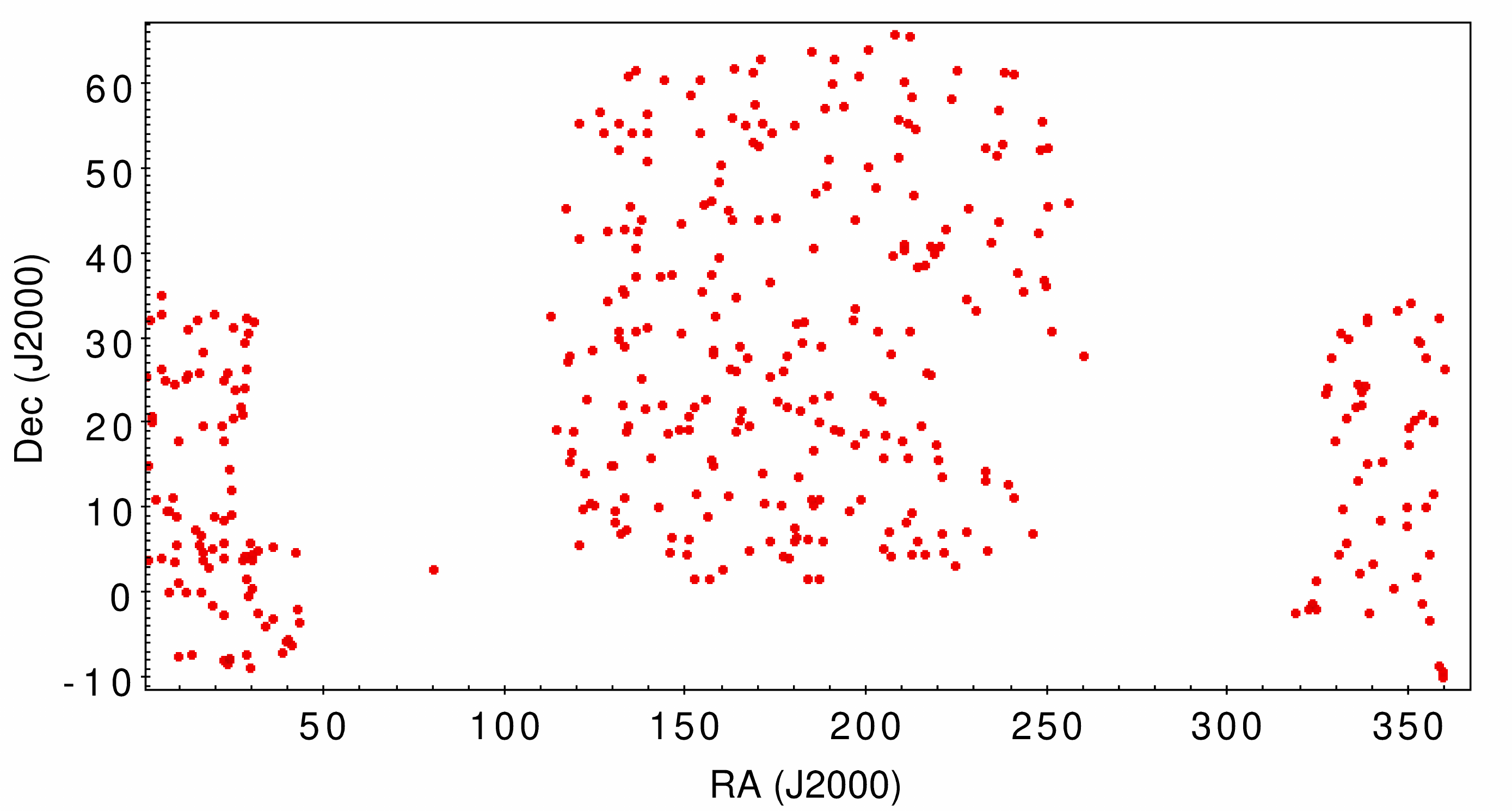}
        \caption{The distribution of candidates on the sky. We can conclude that the distribution is uniform, with no significant bias across the sky.}\label{fig:ra_dec}
    \end{figure}

\begin{figure}
        \centering
        \includegraphics[width=\columnwidth]{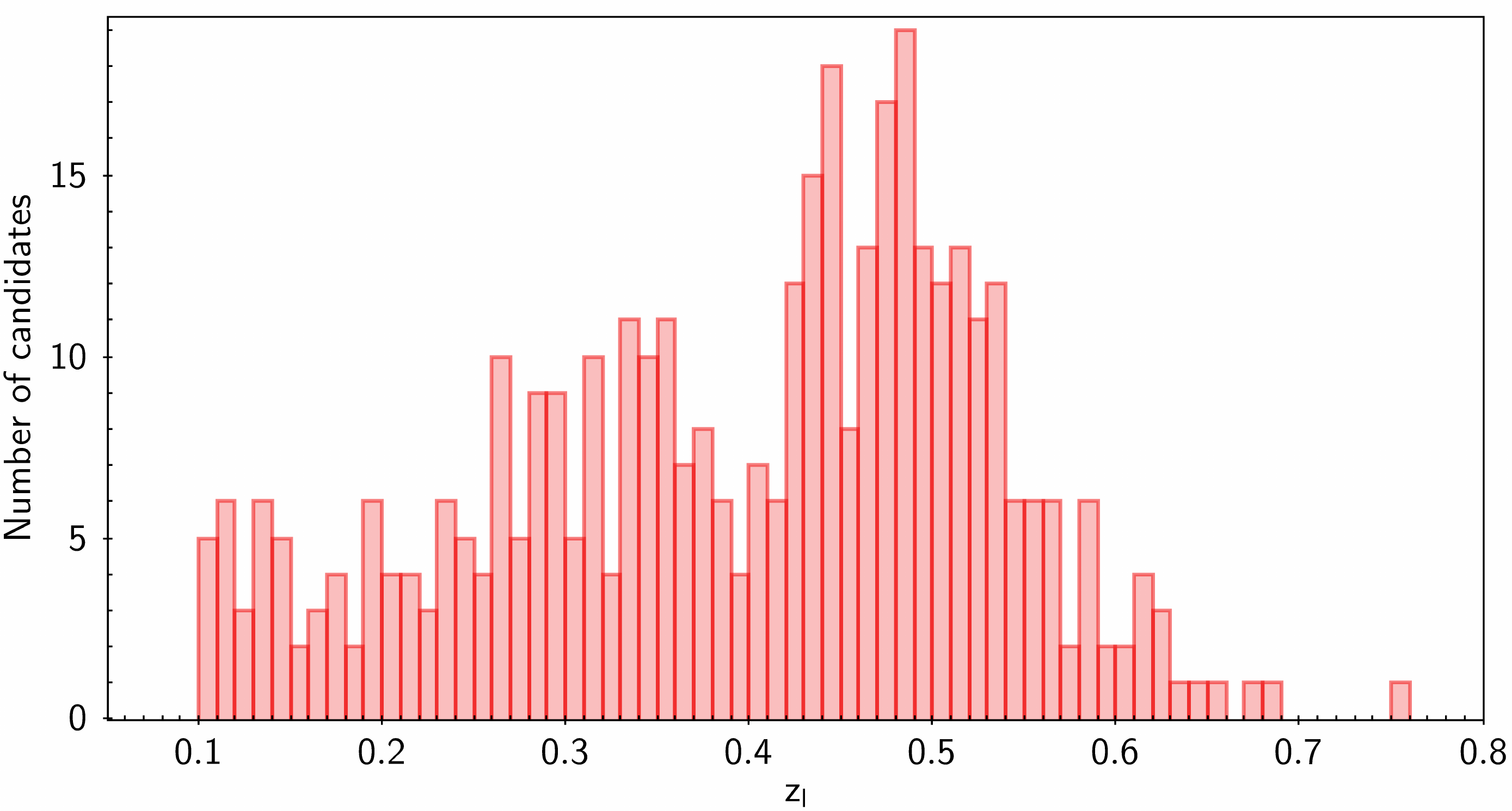}
        \caption{The distribution of lens galaxy redshift of the candidates.}\label{fig:zl}
    \end{figure}

\begin{figure}
        \centering
        \includegraphics[width=\columnwidth]{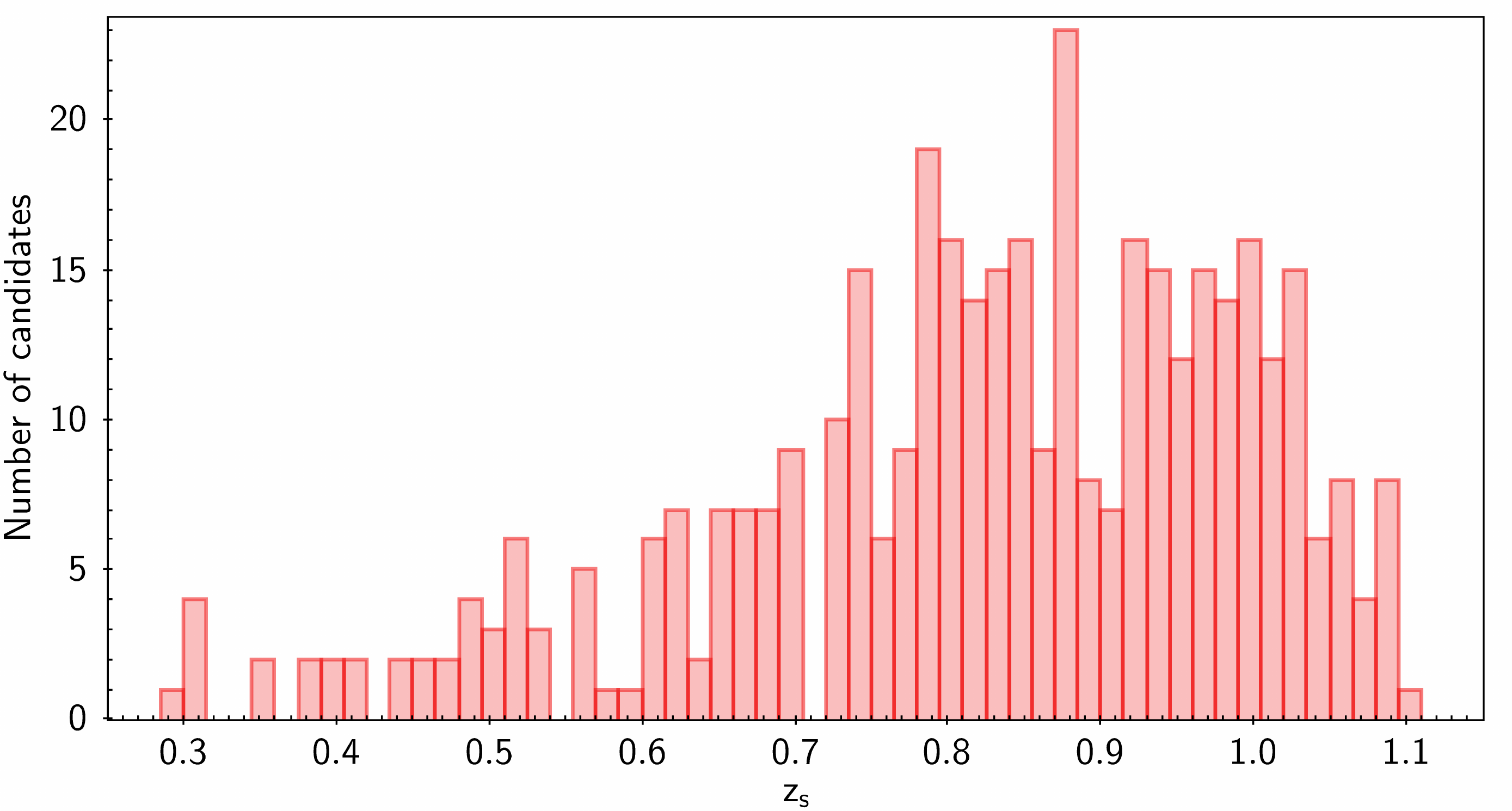}
        \caption{The distribution of source galaxy redshift of the candidates.}\label{fig:zs}
    \end{figure}

    The distribution of the position on the sky, lens galaxy redshift ($z_l$) and source galaxy redshift ($z_s$) of candidates are shown in \autoref{fig:ra_dec}, \autoref{fig:zl} and \autoref{fig:zs}, respectively. We can conclude that there is no significant bias across the sky. Cutouts from existing surveys (DECals, CFHTLS, and HSC) have also been acquired and examined, but none is capable of reaching a definitive conclusion that whether they are strong lensing systems or not.
    
    Here we compiled a list of highly possible candidates catalog in Appendix A.

    \section{Follow-up Observation with CFHT Megacam}\label{sec:cfht}
    All candidates have been checked to see if there is any existing HST coverage. However, none of them has available HST images.
    The final confirmation of strong lensing system requires imaging data from telescopes. Therefore, we applied CFHT Megacam telescope observation time through Telescope Access Program (TAP). Canada-France-Hawaii Telescope (CFHT) is a 3.6m telescope located at Mauna Kea, with average seeing of $\sim 0.6 ''$. Megacam is the wide-field optical imaging facility at CFHT, with 36 CCDs covering 1x1 square degree field-of-view.

    \subsection{Observation details}
\begin{figure}
        \centering
        \includegraphics[width=\textwidth]{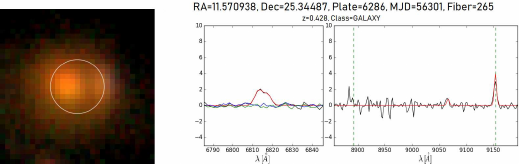}
        \caption{SDSS J004617.02+252041.5. The left panels display false color (gri) images from CFHT Megacam. The white circles represent BOSS fiber, with diameter of 2''. The middle panels are the detected OII doublet in the residual spectra. The black lines indicate the flux in the current fiber, and the blue and green ones refer to flux in adjacent fibers. The red lines indicate best fit of the detected OII doublet. The right panels are part of the residual spectra near the inferred wavelength of OIII and H$\beta$ of the source galaxy in observer frame. The wavelength of [H$\beta$]$\lambda$4858 and [OIII]$\lambda$5007 are pointed out with vertical green dash lines, and red lines indicate best fit. The observer frame wavelength of [OIII]$\lambda$4959 is not indicated in the plot, but the emission line was still fitted with a Gaussian.}\label{fig.cfht_01}
    \end{figure}

\begin{figure}
        \centering
        \includegraphics[width=\textwidth]{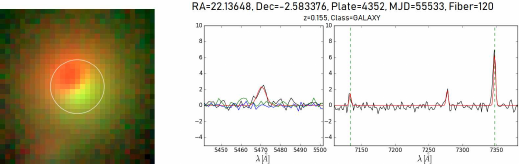}
        \caption{SDSS J012832.31$-$023500.0. See \autoref{fig.cfht_01} for description of the layout of the figure.}\label{fig.cfht_02}
    \end{figure}

\begin{figure}
        \centering
        \includegraphics[width=\textwidth]{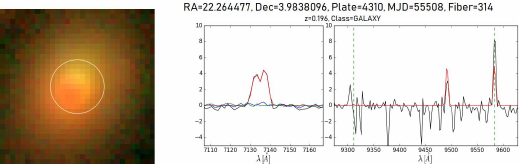}
        \caption{SDSS J012903.47+035901.7. See \autoref{fig.cfht_01} for description of the layout of the figure.}\label{fig.cfht_03}
    \end{figure}

\begin{figure}
        \centering
        \includegraphics[width=\textwidth]{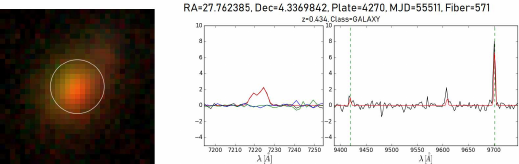}
        \caption{SDSS J015102.97+042013.1. See \autoref{fig.cfht_01} for description of the layout of the figure.}\label{fig.cfht_04}
    \end{figure}

\begin{figure}
        \centering
        \includegraphics[width=\textwidth]{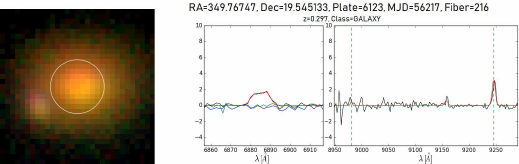}
        \caption{SDSS J231904.19+193242.4. See \autoref{fig.cfht_01} for description of the layout of the figure.}\label{fig.cfht_05}
    \end{figure}

    Due to limited observation time, we only chose the 5 candidates that are visible in September 2018. Since the average seeing of CFHT is $\sim 0.6 ''$, all chosen targets have estimated Einstein radius greater than $0.6''$, with three targets close to $1.0''$ and two targets close to $1.5''$. Since no existing high resolution imaging data is available for these targets for exposure time calculation, the brightness of the distorted source images is estimated with data provided in \cite{bells}, given both our and their targets are selected with SDSS-III BOSS spectra with similar methods. We found that exposure time of 150 seconds is enough to achieve $\text{SNR}=10$. 7 images are taken for each target each band, resulting in a total exposure time of $160\text{ sec}\times7=1120\text{ sec}$ per band per target. Since the seeing might not be good enough to clearly separate the arcs and rings from the lens galaxy, we believe that color images might increase our chance of confirmation since there is often color difference between the lens galaxy and the source galaxy (usually bluer). Therefore, we requested images from three bands: $g$, $r$, and $i$. Their spectra, detected emission lines and CFHT false color (gri) images are shown in \autoref{fig.cfht_01} -- \autoref{fig.cfht_05}. Detailed information about these 5 systems are shown in \autoref{tab.cfht}.
    
    \begin{table}
        \centering
        \caption{Information about 5 targets proposed for CFHT Megacam observation. The first column is the name of the targets, the second column is the plate, modified Julian Date of spectra and the fiber id. The third column, $z_l$, is the redshift of lens galaxy, and the forth column, $z_s$, is the estimated source galaxy redshift. The fifth column, $R_E$, is the calculated Einstein radius, assuming $H_0=70\ \rm km\ s^{-1} Mpc^{-1}$ and $\Omega_m=0.3$. The last column, $\sigma_v$, is the velocity dispersion from BOSS spectra.}\label{tab.cfht}
        \begin{tabular}{c c c c c c}
            \hline\hline
            Name & Plate - MJD - FiberID & $z_{l}$ & $z_{s}$ & $R_E (arcsec)$ & $\sigma_v$ (km/s) \\
            \hline
            SDSS J004617.02+252041.5 & 6286 - 56301 - 265 & 0.428 & 0.829 & 0.841 & 253.54 \\
            SDSS J012832.31$-$023500.0 & 4352 - 55533 - 120 & 0.155 & 0.468 & 1.093 & 245.95 \\
            SDSS J012903.47+035901.7 & 4310 - 55508 - 314 & 0.196 & 0.915 & 1.373 & 248.99 \\
            SDSS J015102.97+042013.1 & 4270 - 55511 - 571 & 0.434 & 0.938 & 0.942 & 271.45 \\
            SDSS J231904.19+193242.4 & 6123 - 56217 - 216 & 0.297 & 0.848 & 1.408 & 281.59 \\
            \hline\hline
        \end{tabular}
    \end{table}
    
    \subsection{CFHT observation results}
    Images from CFHT Megacam are processed by following the steps described below: Star and galaxy Catalog was first generated with SExtractor \citep{1996A&AS..117..393B} on each image and then cross-matched with SDSS DR7 catalog with SCAMP (Software for Calibrating AstroMetry and Photometry, \citet{2006ASPC..351..112B}) for astrometric calibration and photometric calibration. Then all images of the same band of the same target were coadded with SWarp \citep{2002ASPC..281..228B}. Cutouts were made and combined into false color (gri) images for visual inspection, as illustrated from \autoref{fig.cfht_01} to \autoref{fig.cfht_05}. Each cut out is 5.58x5.58 $\rm arcsec^2$.

    From these images, we can conclude that SDSS J012832.31$-$023500.0, SDSS J012903.47+035901.7 and SDSS J015102.97+042013.1 are not strong lensing systems, and SDSS J004617.02+252041.5 or SDSS J231904.19+193242.4 are strong lensing candidates. In the images of these two targets, we clearly detect one distorted source galaxy image, but their counter images are probably out-shined by the central galaxy. Further investigation will be discussed in the next section.

    \section{Discussion}\label{sec:dis}
    As discussed in Section \ref{sec:cfht}, due to limited seeing, none of the 5 systems have an image with clear separation. However, for 2 of them, SDSS J004617.02+252041.5 and SDSS J231904.19+193242.4, we capture one clearly separated source image, which could reveal more information about the system and provide us with a reasonable estimation of the position of the counter-image. Using GALFIT \citep{2010AJ....139.2097P}, we successfully removed the lens galaxies for these two systems and obtained the position of the visible source image with fitting. With a simple Singular Isothermal Sphere (SIS) model, we estimate the separation of their counter-image to their respective central galaxy, shown in \autoref{tab.lens_info}. Each pixel in the images corresponds to $0.187''$ on the sky. Since the distorted source images are only slightly larger than ten pixels, the magnitude estimation by GALFIT has very large errors and are not included.
    \begin{table}
        \centering
        \caption{Measured and estimated information of the two systems. The second row refers to the pixel position of the center of the lens galaxy, obtained with GALFIT. The third row indicates the pixel position of the center of the visible distorted source image, also obtained with GALFIT. The last row $R_2$ refers to the estimated separation of the undetected counter image from the lens galaxy with the SIS model.}
        \label{tab.lens_info}
        \begin{tabular}{c|c c}
            \hline\hline
            Object Name & J004617.02+232041.5 & J231904.19+193242.4 \\
            \hline
            Lens Center (pixel) & 14.304, 16.294 & 16.571, 17.210 \\
            Image Center (pixel) & 21.771, 15.512 & 8.297, 13.117 \\
            $R_E$ (arcsec) & 0.841 & 1.408 \\
            $R_2$ (arcsec) & 0.300 & 1.089 \\
            \hline\hline
        \end{tabular}
    \end{table}
    Given the average seeing of these images are 0.55'', it is safe to conclude that the lens galaxy could easily outshine the counter-image, although with better seeing condition, such as HST or Keck, it would be easy to detect distorted counter-image of the source galaxy.
    
     \cite{bells} published a catalog of 36 confirmed strong lensing candidates confirmed by HST. We cross checked our results with their list, and we recovered 30 of theirs. The other seven are lost since we are more conservative in our selection algorithm. A number of candidates inside the BELLS catalog have [OII]$\lambda$3727, 3729 doublet beyond 9200$\AA$ in the observer frame, which are discarded in our selection algorithm for lowering overall contamination rate. Since the final confirmation of strong lensing system always requires long exposure time from best telescopes, we consider the 80\% recovery rate as acceptable.
     
     The ratio of discovered strong galaxy-galaxy lensing system candidates discovered by the algorithm is $0.029\%$, which is considerably lower than the theoretical estimation by \citet{2008ApJ...685...57D} who gives a value of $0.5\%$ to $1.3\%$, and the lensing probability in SLACS \citep{slacs}, which is about $0.1\%$. However, this value is comparable to the ratio of BELLS, which is $36/133852=0.027\%$ \citep{bells}. The reasons of this lowered lensing probability compared to previous searches and theoretical estimation are two-folds. Firstly, the diameter of the fiber is reduced from $3''$ to $2''$, and according to \citet{2008ApJ...685...57D}, the selection effect introduced by finite fiber size plays an important role in limiting the maximum number of strong galaxy-galaxy lensing system that could be discovered. Secondly, we only considered those candidates with more than one emission line and with a large enough estimated Einstein radius, which would also remove many of the potential strong galaxy-galaxy lensing systems.

    \section{Conclusion}\label{sec:con}
    In this work, we applied similar selection algorithm to the entire SDSS-III BOSS galaxies and were able to automate many steps that were used to be manually done in \cite{bells}. During this process we also generated a list of 100 highly possible candidates. Follow-up CFHT observations were preformed on five targets, with two of them being potential strong lensing systems. Although only one distorted image can be seen and its counter images probably outshined by central galaxy, we are able to confirm that they are two galaxies detected at different redshift. Assuming these are strong lensing systems, we are capable to calculate the position of the source image closer to the central galaxy and they could not be clearly resolved under the seeing with CFHT. Further follow-up with telescopes with higher resolution power, such as Keck or HST, might be able to clearly reveal the true nature of these systems.
    
    Faced with limited seeing, one can extend the list of confirmation of candidates with Hubble Space Telescope and Keck AO.  As indicated in \citet{bells}, the complete BOSS galaxy sample is estimated to produce at least a few hundred strong lenses, which, if confirmed, would greatly increase the amount of galaxy-galaxy strong lensing systems known. With a larger number and higher quality imaging data, we plan to apply more complicated lens modeling, such as SIE Lens Models with Sersic source models described in \citet{slacs} and \citet{bells}, to shed light on the evolution of the relationship between stellar mass, luminosity and galaxy structure.

    \begin{acknowledgements}
    The author appreciate the anonymous referee for his/her helpful suggestions which improved the paper. The author thanks Drs. Shude Mao, Yiping Shu, Dandan Xu and Mr. Yunchong Wang for helpful discussions and contributions to this project, which is partly supported by the National Key Basic Research and Development Program of China (No. 2018YFA0404501 to SM), by the National Science Foundation of China (Grant No. 11821303 and 11761131004 to SM).
    
    This research uses data obtained through the Telescope Access Program (TAP), which has been funded by the National Astronomical Observatories, Chinese Academy of Sciences, and the Special Fund for Astronomy from the Ministry of Finance. Based on observations obtained with MegaPrime/MegaCam, a joint project of CFHT and CEA/DAPNIA, at the Canada-France-Hawaii Telescope (CFHT) which is operated by the National Research Council (NRC) of Canada, the Institut National des Science de l'Univers of the Centre National de la Recherche Scientifique (CNRS) of France, and the University of Hawaii. 
    
    This paper made use of SDSS-III data. Funding for SDSS-III has been provided by the Alfred P. Sloan Foundation, the Participating Institutions, the National Science Foundation, and the U.S. Department of Energy Office of Science. The SDSS-III web site is http://www.sdss3.org/.

    SDSS-III is managed by the Astrophysical Research Consortium for the Participating Institutions of the SDSS-III Collaboration including the University of Arizona, the Brazilian Participation Group, Brookhaven National Laboratory, Carnegie Mellon University, University of Florida, the French Participation Group, the German Participation Group, Harvard University, the Instituto de Astrofisica de Canarias, the Michigan State/Notre Dame/JINA Participation Group, Johns Hopkins University, Lawrence Berkeley National Laboratory, Max Planck Institute for Astrophysics, Max Planck Institute for Extraterrestrial Physics, New Mexico State University, New York University, Ohio State University, Pennsylvania State University, University of Portsmouth, Princeton University, the Spanish Participation Group, University of Tokyo, University of Utah, Vanderbilt University, University of Virginia, University of Washington, and Yale University. 
    \end{acknowledgements}
    
    \bibliographystyle{raa}
    \bibliography{msRAA20190151R1}{}

    \newpage
    \appendix
    \section{Highly Probable Candidate List}
    \begin{table}[h]
        \centering
        \caption{A list of top-100 candidates from the selection algorithm. The first and second column are the Right Ascension (RA) and Declination (Dec) in J2000, respectively. The third column is the plate number (Plate), modified Julian date of the spectra observation (MJD), and fiber id of the target (FiberID). The fourth column, $z_{l}$, is the redshift of the lens galaxy, which is calculated by the BOSS pipeline. The fifth column, $z_{s}$, is the estimated redshift of the source galaxy, assuming the detected peak is the OII doublet. The sixth column is the velocity dispersion of the lens galaxy measured by BOSS pipeline using the width of emission lines. The seventh column, $R_E$ (''), refers to the estimated Einstein radius with SIS model in arc-seconds, assuming $H_0=70\ \rm km\ s^{-1}Mpc^{-1}$ and $\Omega_m=0.3$. The eighth column, $\rm I$, is the i-band magnitude of lens galaxy measured by SDSS photometry with extinction correction applied.}\label{tab:appdx}
        \begin{tabular}{c c c c c c c c c c c c}
            \hline\hline
            RA (J2000) & Dec (J2000) & Plate-MJD-FiberID & $z_l$ & $z_s$ & $\sigma_v$ (km/s)& $R_E$ ('') & I \\
            \hline
  09:17:26.81 & +31:09:25.1 & 5808 - 56325 - 0434 & 0.239 & 0.821 & 221.83 & 0.945 & 19.158\\
  11:34:48.88 & +54:11:13.1 & 6697 - 56419 - 0386 & 0.302 & 0.614 & 276.51 & 1.036 & 19.547\\
  15:38:29.13 & +41:17:54.3 & 6050 - 56089 - 0379 & 0.587 & 0.931 & 329.13 & 0.982 & 20.857\\
  10:02:48.66 & +20:48:52.0 & 5784 - 56029 - 0104 & 0.372 & 0.494 & 290.96 & 0.550 & 20.138\\
  10:09:12.73 & +01:31:42.7 & 4738 - 55650 - 0294 & 0.479 & 0.947 & 235.08 & 0.690 & 20.678\\
  17:19:19.83 & +27:53:16.4 & 5003 - 55715 - 0681 & 0.530 & 0.952 & 227.77 & 0.572 & 20.563\\
  01:29:03.47 & +03:59:01.7 & 4310 - 55508 - 0314 & 0.196 & 0.915 & 252.72 & 1.373 & 18.811\\
  23:47:54.15 & +20:03:08.1 & 6126 - 56269 - 0845 & 0.411 & 0.743 & 212.89 & 0.523 & 20.776\\
  11:32:12.17 & +06:07:12.7 & 4767 - 55946 - 0563 & 0.557 & 0.856 & 293.82 & 0.746 & 21.129\\
  13:40:58.11 & +18:37:52.4 & 5862 - 56045 - 0110 & 0.244 & 0.646 & 203.35 & 0.698 & 19.221\\
  11:45:59.88 & +10:22:41.1 & 5380 - 55980 - 0804 & 0.385 & 0.916 & 241.04 & 0.874 & 20.365\\
  12:21:36.90 & +16:52:39.3 & 5849 - 56033 - 0068 & 0.437 & 0.975 & 246.95 & 0.858 & 20.082\\
  01:28:32.32 & -02:35:00.0 & 4352 - 55533 - 0120 & 0.155 & 0.468 & 242.40 & 1.093 & 18.556\\
  13:45:24.20 & +07:05:46.6 & 4865 - 55713 - 0836 & 0.229 & 0.628 & 209.69 & 0.761 & 19.038\\
  01:51:02.97 & +04:20:13.1 & 4270 - 55511 - 0571 & 0.434 & 0.938 & 261.58 & 0.942 & 20.918\\
  00:47:51.06 & +31:04:32.1 & 6872 - 56540 - 0443 & 0.643 & 0.848 & 447.56 & 1.170 & 20.958\\
  09:03:42.53 & +40:32:46.2 & 4605 - 55971 - 0765 & 0.437 & 0.529 & 390.85 & 0.685 & 20.810\\
  23:07:58.78 & +33:13:19.0 & 6504 - 56540 - 0833 & 0.569 & 1.037 & 527.48 & 3.084 & 20.631\\
  22:22:19.79 & +21:56:58.4 & 5946 - 56101 - 0884 & 0.487 & 0.952 & 262.30 & 0.846 & 20.777\\
  14:12:17.55 & +46:56:30.7 & 6751 - 56368 - 0355 & 0.402 & 0.654 & 223.88 & 0.501 & 20.396\\
  14:15:02.80 & +54:42:37.5 & 6797 - 56426 - 0149 & 0.618 & 0.786 & 321.40 & 0.540 & 20.560\\
  23:19:04.19 & +19:32:42.5 & 6123 - 56217 - 0216 & 0.297 & 0.848 & 285.39 & 1.408 & 19.461\\
  23:35:45.29 & -01:15:17.5 & 4357 - 55829 - 0561 & 0.486 & 0.881 & 236.25 & 0.631 & 20.492\\
  13:46:47.26 & +04:10:19.5 & 4785 - 55659 - 0610 & 0.414 & 0.624 & 450.30 & 1.769 & 20.168\\
  08:54:26.52 & +07:27:01.7 & 4867 - 55924 - 0865 & 0.461 & 0.768 & 351.37 & 1.257 & 21.148\\
  10:27:22.44 & +46:06:47.5 & 6659 - 56607 - 0299 & 0.521 & 0.751 & 298.71 & 0.684 & 21.133\\
  08:57:12.83 & +19:43:52.1 & 5175 - 55955 - 0804 & 0.377 & 0.628 & 234.19 & 0.573 & 19.821\\
  13:30:57.19 & +47:39:54.2 & 6743 - 56385 - 0086 & 0.349 & 0.664 & 236.65 & 0.698 & 19.568\\
  00:19:34.83 & +32:45:19.7 & 7130 - 56568 - 0432 & 0.114 & 0.648 & 213.57 & 1.054 & 17.684\\
  02:00:13.95 & +03:53:20.1 & 4270 - 55511 - 0007 & 0.429 & 0.754 & 444.35 & 2.191 & 20.983\\
  11:54:49.31 & +04:05:41.8 & 4765 - 55674 - 0115 & 0.443 & 0.649 & 294.56 & 0.706 & 20.555\\
  01:46:48.16 & +22:00:09.6 & 5108 - 55888 - 0811 & 0.267 & 0.745 & 214.99 & 0.798 & 19.297\\
\hline\hline
        \end{tabular}
    \end{table}
  \begin{table}
        \centering
        \caption{\autoref{tab:appdx} - continued}\label{tab:appdx2}
        \begin{tabular}{c c c c c c c c c c c c}
            \hline\hline
            RA (J2000) & Dec (J2000) & Plate-MJD-FiberID & $z_l$ & $z_s$ & $\sigma_v$ (km/s)& $R_E$ ('') & I \\
            \hline
  09:33:14.20 & +22:05:16.6 & 5789 - 56246 - 0362 & 0.350 & 0.695 & 216.82 & 0.615 & 20.065\\
  14:27:09.14 & +25:58:43.5 & 6015 - 56096 - 0927 & 0.356 & 0.756 & 279.67 & 1.086 & 20.907\\
  12:00:51.72 & +06:03:13.0 & 4847 - 55931 - 0040 & 0.334 & 0.738 & 330.80 & 1.579 & 19.541\\
  14:09:49.98 & +58:22:41.7 & 6804 - 56447 - 0957 & 0.360 & 0.744 & 285.14 & 1.100 & 19.297\\
  07:29:51.26 & +32:37:17.3 & 4444 - 55538 - 0596 & 0.342 & 0.839 & 270.38 & 1.140 & 19.394\\
  00:46:17.03 & +25:20:41.5 & 6286 - 56301 - 0265 & 0.428 & 0.829 & 260.19 & 0.841 & 20.216\\
  23:53:33.02 & -08:38:58.2 & 7166 - 56602 - 0784 & 0.167 & 0.657 & 209.51 & 0.904 & 18.061\\
  00:13:10.41 & +11:03:52.9 & 6113 - 56219 - 0841 & 0.578 & 0.811 & 312.92 & 0.693 & 20.873\\
  22:10:54.47 & +05:47:40.4 & 4427 - 56107 - 0478 & 0.177 & 0.778 & 181.41 & 0.700 & 18.946\\
  01:00:02.69 & +32:05:19.9 & 6593 - 56270 - 0403 & 0.487 & 0.875 & 241.41 & 0.651 & 20.169\\
  13:22:18.78 & +64:00:22.6 & 6822 - 56711 - 0571 & 0.482 & 0.786 & 247.31 & 0.599 & 20.625\\
  23:47:36.70 & +20:23:01.9 & 6126 - 56269 - 0834 & 0.450 & 0.588 & 304.53 & 0.558 & 19.960\\
  02:05:56.30 & -02:20:02.1 & 4347 - 55830 - 0156 & 0.283 & 0.462 & 256.85 & 0.687 & 19.056\\
  13:07:22.99 & +33:22:53.0 & 6488 - 56364 - 0927 & 0.589 & 1.022 & 382.08 & 1.510 & 21.068\\
  23:22:25.23 & +34:14:58.3 & 7139 - 56568 - 0687 & 0.292 & 0.613 & 272.90 & 1.043 & 19.265\\
  15:50:22.42 & +52:49:35.8 & 6715 - 56449 - 0938 & 0.442 & 0.734 & 241.26 & 0.594 & 20.787\\
  00:27:32.70 & +09:37:17.4 & 6195 - 56220 - 0345 & 0.463 & 0.864 & 209.84 & 0.520 & 20.571\\
  01:39:28.96 & +20:38:23.4 & 7242 - 56628 - 0051 & 0.597 & 0.879 & 370.07 & 1.076 & 21.022\\
  01:25:59.26 & +19:36:41.7 & 5135 - 55862 - 0850 & 0.101 & 0.585 & 226.37 & 1.191 & 17.946\\
  12:38:13.00 & +23:15:37.6 & 5983 - 56310 - 0025 & 0.480 & 0.852 & 246.12 & 0.668 & 20.341\\
  12:38:46.67 & +51:00:56.8 & 6674 - 56416 - 0262 & 0.388 & 0.766 & 236.44 & 0.716 & 19.914\\
  10:51:50.40 & +44:01:42.5 & 4689 - 55656 - 0573 & 0.435 & 1.008 & 325.06 & 1.533 & 20.280\\
  12:34:18.66 & +57:08:52.9 & 6832 - 56426 - 0591 & 0.432 & 0.799 & 241.03 & 0.685 & 20.253\\
  09:16:28.33 & +56:30:45.2 & 5725 - 56625 - 0595 & 0.434 & 0.895 & 238.81 & 0.752 & 20.396\\
  08:50:21.70 & +35:47:50.8 & 4602 - 55644 - 0603 & 0.475 & 0.664 & 446.82 & 1.445 & 20.697\\
  23:28:48.60 & +01:48:25.3 & 4284 - 55863 - 0107 & 0.437 & 0.855 & 259.83 & 0.845 & 20.846\\
  23:38:28.85 & +27:41:25.3 & 6516 - 56571 - 0543 & 0.549 & 0.849 & 264.13 & 0.611 & 20.779\\
  01:57:48.01 & -08:51:06.6 & 7184 - 56629 - 0855 & 0.501 & 1.035 & 227.24 & 0.667 & 20.534\\
  09:18:07.92 & +50:49:36.2 & 5730 - 56607 - 0351 & 0.581 & 0.913 & 353.94 & 1.117 & 20.230\\
  13:06:40.98 & +43:51:21.8 & 6621 - 56366 - 0263 & 0.482 & 0.720 & 266.59 & 0.595 & 20.713\\
  16:39:56.53 & +45:34:34.3 & 6027 - 56103 - 0388 & 0.422 & 0.840 & 324.38 & 1.347 & 20.938\\
  13:36:35.91 & +22:33:41.8 & 5998 - 56087 - 0336 & 0.269 & 0.834 & 197.44 & 0.709 & 19.288\\
  10:20:52.84 & +45:44:32.4 & 7385 - 56710 - 0198 & 0.198 & 0.524 & 653.74 & 7.308 & 20.281\\
  14:43:54.17 & +07:02:44.5 & 4858 - 55686 - 0601 & 0.514 & 0.997 & 209.95 & 0.533 & 20.427\\
  15:09:41.98 & +34:32:02.3 & 4717 - 55742 - 0121 & 0.390 & 0.769 & 647.44 & 5.359 & 19.902\\
  00:35:34.70 & +09:03:59.4 & 4540 - 55863 - 0069 & 0.214 & 0.805 & 215.77 & 0.932 & 19.288\\
  02:52:34.60 & -03:31:12.4 & 7054 - 56575 - 0701 & 0.412 & 0.787 & 231.24 & 0.657 & 20.420\\
  23:48:17.74 & +11:39:09.0 & 6150 - 56187 - 0089 & 0.472 & 0.775 & 236.82 & 0.556 & 20.354\\
  10:16:37.24 & +54:18:16.9 & 6696 - 56398 - 0487 & 0.487 & 0.863 & 289.27 & 0.921 & 20.452\\
  13:17:19.45 & +18:40:50.1 & 5867 - 56034 - 0780 & 0.291 & 0.967 & 226.74 & 0.957 & 19.267\\
  00:27:25.59 & +00:06:40.5 & 4220 - 55447 - 0737 & 0.218 & 0.535 & 206.72 & 0.691 & 21.118\\
  10:11:55.93 & +11:33:14.8 & 5331 - 55976 - 0910 & 0.387 & 0.876 & 296.83 & 1.278 & 20.189\\
  21:58:05.89 & +17:56:26.1 & 5020 - 55852 - 0899 & 0.628 & 0.854 & 338.48 & 0.735 & 20.716\\
  \hline\hline
        \end{tabular}
    \end{table}
\begin{table}
        \centering
        \caption{\autoref{tab:appdx2} - continued}
        \begin{tabular}{c c c c c c c c c c c c}
            \hline\hline
            RA (J2000) & Dec (J2000) & Plate-MJD-FiberID & $z_l$ & $z_s$ & $\sigma_v$ (km/s)& $R_E$ ('') & I \\
            \hline
  13:38:47.40 & +05:06:01.8 & 4786 - 55651 - 0678 & 0.271 & 0.560 & 259.91 & 0.938 & 19.113\\
  23:58:12.87 & -09:18:25.0 & 7167 - 56604 - 0642 & 0.376 & 0.468 & 747.56 & 2.886 & 19.817\\
  00:26:02.88 & +09:44:17.2 & 6195 - 56220 - 0424 & 0.246 & 0.389 & 266.92 & 0.714 & 19.237\\
  14:01:07.02 & +41:03:44.0 & 6630 - 56358 - 0844 & 0.448 & 0.870 & 300.27 & 1.115 & 20.661\\
  02:00:58.79 & +04:35:46.1 & 4268 - 55483 - 0644 & 0.469 & 0.758 & 243.72 & 0.576 & 20.419\\
  10:31:25.35 & +32:42:14.2 & 6450 - 56331 - 0081 & 0.610 & 1.008 & 240.59 & 0.554 & 20.927\\
  12:50:21.74 & +19:05:05.4 & 5856 - 56090 - 0868 & 0.500 & 0.822 & 347.36 & 1.191 & 20.418\\
  15:56:18.83 & +12:46:51.6 & 4901 - 55711 - 0573 & 0.552 & 0.944 & 478.26 & 2.348 & 20.915\\
  08:29:19.60 & +54:11:57.4 & 5156 - 55925 - 0574 & 0.361 & 0.676 & 235.53 & 0.678 & 20.510\\
  08:11:14.08 & +22:43:19.7 & 4469 - 55863 - 0216 & 0.165 & 0.918 & 222.58 & 1.122 & 18.488\\
  09:11:34.76 & +25:17:08.9 & 5778 - 56328 - 0079 & 0.566 & 0.794 & 308.63 & 0.678 & 20.589\\
  09:44:16.65 & +06:27:43.5 & 4872 - 55944 - 0061 & 0.338 & 0.867 & 232.94 & 0.872 & 21.344\\
  23:31:45.76 & +29:21:48.7 & 6581 - 56540 - 0976 & 0.313 & 0.679 & 274.55 & 1.078 & 19.867\\
  01:18:01.04 & +08:53:17.4 & 4555 - 56189 - 0210 & 0.359 & 0.801 & 330.19 & 1.575 & 19.591\\
  01:15:48.47 & +05:09:48.8 & 4425 - 55864 - 0376 & 0.337 & 0.607 & 217.98 & 0.558 & 19.852\\
  07:49:21.61 & +27:20:01.0 & 4459 - 55533 - 0605 & 0.492 & 0.682 & 375.22 & 0.993 & 20.660\\
  11:59:15.50 & +55:09:07.2 & 6839 - 56425 - 0117 & 0.489 & 1.025 & 270.56 & 0.963 & 20.245\\
  09:55:05.21 & +30:32:30.6 & 5800 - 56279 - 0591 & 0.473 & 1.013 & 225.59 & 0.685 & 20.351\\
  08:51:22.62 & +11:09:29.2 & 5291 - 55947 - 0681 & 0.503 & 0.838 & 452.24 & 2.050 & 21.057\\
  12:21:43.03 & +40:43:42.5 & 6633 - 56369 - 0164 & 0.216 & 0.501 & 246.50 & 0.944 & 18.726\\
  15:47:10.74 & +56:52:52.0 & 6791 - 56429 - 0025 & 0.441 & 0.993 & 247.26 & 0.867 & 20.043\\
  11:48:17.87 & +26:06:28.9 & 6411 - 56331 - 0427 & 0.291 & 0.743 & 180.38 & 0.529 & 19.664\\
  13:48:40.76 & +39:40:45.4 & 4710 - 55707 - 0601 & 0.362 & 0.699 & 324.13 & 1.326 & 20.860\\
  08:02:00.13 & +55:17:23.8 & 5945 - 56213 - 0636 & 0.461 & 0.999 & 273.23 & 1.020 & 20.426\\
  12:08:21.58 & +29:33:32.1 & 6472 - 56362 - 0375 & 0.447 & 0.876 & 232.59 & 0.676 & 20.647\\
            \hline\hline
        \end{tabular}
    \end{table}
\label{lastpage}
\end{document}